\documentstyle[12pt]{article}
\title{\bf Dissipative particle dynamics with energy conservation}
\author{J.Bonet Avalos and A.D. Mackie\\
Departament d'Enginyeria
Qu\'{\i}mica, ETSEQ, 
Universitat Rovira i Virgili.\\ Carretera de Salou s/n, 43006
Tarragona (Spain)}
\begin{document}
\maketitle\parskip 2ex
\renewcommand{\theequation}{\arabic{equation}}

\begin{abstract}
The stochastic differential equations for a model of dissipative particle
dynamics with both total energy and total momentum conservation in the
particle-particle interactions are presented. The corresponding
Fokker-Planck equation for the evolution of the probability distribution
for the system is deduced together with the corresponding
fluctuation-dissipation theorems ensuring that the {\em ab initio} chosen
equilibrium probability distribution for the relevant variables is a
stationary solution. When energy conservation is included, the system can
sustain temperature gradients and heat flow can be modeled.
\end{abstract}


\setcounter{equation}{0}        

        Much attention has recently been paid to the simulation strategy
for the dynamics of complex fluids known as {\em Dissipative
Particle Dynamics} (DPD), which was first introduced by Hoogerbrugge {\em
et al.}\cite{Hoo}. In this model, a system of {\em mesoscopic} particles
can interact via direct conservative potentials, as in molecular
dynamics (MD) simulations but, in addition, the particles exert 
friction and brownian forces on each other. The dissipative and
random interactions are chosen in such a way that the center of mass
motion of each interacting pair is insensitive to their action.  In this
way, the system relaxes to its equilibrium much
faster than in MD simulations and, at the same time, the interaction
conserves the total momentum. This second feature allows the system to
exhibit a hydrodynamic behaviour from a macroscopic point of view. The
model is isotropic and Galilean-invariant, in contrast with models defined
on a lattice, and has the potential to be computationally efficient.

        Thus, DPD appears as an interesting tool although its capability
for making quantitative predictions on the dynamics of complex systems
remains still to be explored. Some steps forward, however, have already
been taken. For instance, Espa\~nol {\em et
al.}\cite{Esp1} have derived the fluctuation-disipation theorem
appropriate to 
these dissipative particles, 
the true hydrodynamic behaviour of the model has been
proved\cite{Esp2} and the relationship of the macroscopic transport
coefficients with the mesoscopic parameters defining the model has been
established\cite{Ern}. The DPD method has also been applied with remarkable
success to account for the many-body problem of the hydrodynamic
interactions in colloidal suspensions\cite{Lek} as well as to domain
growth in binary immiscible fluids\cite{Cov}.

        As originally formulated, DPD can only deal with isothermal
conditions since the energy is not conserved in the interaction\cite{Ern}.
The system therefore cannot sustain temperature
gradients and, hence, no heat flow can be
modeled\cite{Ern}. This limitation excludes the DPD simulation strategy
from those problems in which non-equilibrium situations involving
temperature gradients or heat generation are important, as is the
case for convection or in chemical reactions. The aim of
this letter is to extend the DPD algorithm to account for both momentum and
energy conservation in the particle-particle interaction, whilst
maintaining the irreversible nature of these interactions.

        The dissipative particles can be viewed as {\em lumps}\cite{Ern} or
{\em clusters} of true physical particles or as particles with internal
structure bearing some degrees of freedom. The DPD model is mesoscopic in
nature since it resolves only the overall center-of-mass
motion of the cluster and 
avoids the description of the variables specifying its {\em internal
state}. We will account for the energy stored in
the internal degrees of freedom of the particles without explicit
consideration of any internal Hamiltonian, in a model inspired by the
treatment of hydrodynamic fluctuations\cite{Maz1,vSa,Lan}.

        Our model is based on the following assumptions:  
\begin{enumerate}
\item The system contains $N$ particles interacting with each other via
conservative as well as dissipative interactions. The conservative
interactions are described by the Hamiltonian
\begin{equation}
H(\{\vec{r}_i\},\{\vec{p}_i\}) \equiv \sum_i^N \left\{ \frac{p_i^2}{2m}
+\sum_{j>i}^N \psi (r_{ij}) +\psi^{ext}(\vec{r}_i) \right\}          \label{1}
\end{equation}
where $r_{ij} \equiv |\vec{r}_i-\vec{r}_j|$. The Hamiltonian depends on
the momenta and positions of all the particles. 
The particles interact with each other through pair potentials, $\psi
(r_{ij})$, which depend only on the distance between the particles, and
with an external field, $\psi^{ext}(\vec{r}_i)$.
\item In addition, the particles can store energy in some internal degrees
of freedom. The internal energy, $u_i$ with $u_i \geq 0$, is introduced as a
new relevant coordinate. Together with the momentum $\vec{p}_i$ and the
position $\vec{r}_i$ this completely specifies the state of the dissipative
particle at a given instant $t$.
\item The particle-particle interaction is pairwise and conserves the
total momentum, the total angular momentum and the total energy when the
internal energy of the pair is taken into account. 
\item The internal states of the particle have no dynamics in the sense
that they are always in equilibrium with themselves. This allows us to
define a function $s_i(u_i)$. This
function can arbitrarily be chosen according to the user's need, only
constrained to the requirements: a)  $s_i$ must be a
differentiable monotonously increasing function of $u_i$, so that the
function $u_i(s_i)$ exists and $\theta_i \equiv 
\partial u_i/\partial s_i$ exists and is always positive, and b)
$s_i(u_i)$ is a {\em concave} function of its argument\cite{Ca}.
Defined in this way, $s_i$ can be viewed as a mesoscopic
{\em entropy} of the $i^{th}$ particle, and $\theta_i$ can be seen as
the particle's {\em temperature}. The change in
$u_i$ and in $s_i$ are related by a Gibbs equation\cite{Maz1} $\theta_i \,
ds_i = du_i$, which implies $\theta_i \, \dot{s_i} = \dot{u_i}$,
where the dot over the variables is used to denote time-differentiation
from now on.
\item In the absence of random terms, the pairwise particle-particle
interaction is irreversible and must satisfy $\dot{s_i} + \dot{s_j} \geq
0$, where $i$ and $j$ label an arbitrary pair of particles in interaction
with each other.
\item $u_i$, $s_i$ and $\theta_i$ must
remain unchanged under a Galilean transformation, so that these variables
are true scalars.
\item The equilibrium probability distribution
for the relevant variables of the system under isothermal conditions is
chosen to be 
\begin{equation}
P_e(\{\vec{r}_i\},\{\vec{p}_i\},\{u_i\}) \sim
e^{-H(\{\vec{r}_i\},\{\vec{p}_i\})/kT}\prod_{i=1}^N e^{s_i(u_i)/k-u_i/kT}
                \label{Pe}
\end{equation}
where $k$ is Boltzmann's constant and $T$ is the thermodynamic
temperature. The first factor on the right hand side of eq. (\ref{Pe})
contains the probability distribution for the variables
$\{\vec{r}_i\},\{\vec{p}_i\}$ as given from equilibrium statistical
mechanics. The second factor on the right hand side corresponds, in turn,
to the 
probability distribution for the internal energy of the particles as
obtained from equilibrium fluctuation theory\cite{Ca}. The maximum of
$s_i(u_i)/k-u_i/kT$ takes place at $\theta_i=T$, 
in agreement with our interpretation of $\theta_i$ as the particle's
temperature. 
Once the equilibrium probability distribution is set, all the
thermodynamic properties of the model can be derived from the 
partition function $Z(T,V,N) \equiv \int (\prod_i d\vec{p}_i d\vec{r}_i du_i)\;
P_e(\{\vec{r}_i\},\{\vec{p}_i\},\{u_i\})$.
\end{enumerate}

        We proceed to the analysis of the dynamics of the model defined so
far, although the details of the calculation will be given elsewhere. For
simplicity, we will analyse two particles, $i$ and $j$ say, due to
the pairwise additivity of the interaction, and give the complete
expression at the end. Since the state of the system is specified by the
set $\{\vec{r}_i\}$, $\{\vec{p}_i\}$ and $\{u_i\}$, we have to supply
equations for the evolution of these variables. The change in the position
and momentum of the $i^{th}$ particle due to the interactions with the
$j^{th}$ particle is given by
\begin{eqnarray}
\dot{\vec{r}}_i &=& \frac{\vec{p}_i}{m}   \label{4}\\
\dot{\vec{p}}_i &=& \vec{F}_{ij}^C+ \vec{F}_i^{ext} + \vec{F}_{ij}^D
+\vec{F}_{ij}^R   \label{5}
\end{eqnarray}
where $\vec{F}_{ij}^C=-\partial \psi(r_{ij})/\partial \vec{r}_i$ and
$\vec{F}_i^{ext} \equiv -\partial \psi^{ext}(\vec{r}_i)/\partial
\vec{r}_i$ are the 
forces due to the conservative interactions. $\vec{F}_{ij}^C$ is, by
construction, directed along the unit vector $\hat{r}_{ij} \equiv
(\vec{r}_j-\vec{r}_i)/r_{ij}$. $\vec{F}_{ij}^D$ stands for the
dissipative particle-particle interaction force and $\vec{F}_{ij}^R$ is
the random force associated with the former. The mechanisms driving the
change in the internal energy of the particles are assumed to be of two
kinds. On the one hand, the work done by the dissipative forces increases
the 
internal energy of the interacting particles. Since they are identical,
we assume that this work is shared in equal amounts by the particles and
is irrespective of the temperatures $\theta_i$ and $\theta_j$. At
the same time, the action of the random force {\em cools} the particles
transfering internal energy back to mechanical energy. We want, in addition,
that viscosity and thermal conductivity can be independently modeled.
Hence, we also consider, on the other hand, that interacting particles can
exchange internal energy if $\theta_i \neq \theta_j$, something that we
can call {\em mesoscopic heat flow} $\dot{q}_{ij}^D$ between particles.
Associated with this dissipative current, we also add a random heat flow
$\dot{q}_{ij}^R$. Thus, we write
\begin{equation}
\dot{u}_i= \frac{1}{2m} (\vec{p}_j-\vec{p}_i) \cdot
\left(\vec{F}_{ij}^D+\vec{F}_{ij}^R \right) +\dot{q}_{ij}^D+\dot{q}_{ij}^R
        \label{6}
\end{equation}
The dissipative and random terms have
to be of the form $\vec{F}_{ij}^{D,R} \sim \hat{r}_{ij}$ and
$\dot{q}_{ij}^{D,R} = -\dot{q}_{ji}^{D,R}$. Otherwise, the requirements of
point 3 would be 
violated. 

        The analysis of point 5, permits the identification of the
so-called {\em thermodynamic forces}\cite{Maz1} causing the dissipative
currents to exist. In our case, it is rather intuitive that the friction
forces are due to the momentum difference between interacting particles,
and that the heat flow is due to a temperature difference. We, thus, assume
linear relationships of the form
\begin{equation}
\vec{F}_{ij}^D =\zeta(r_{ij}) \frac{1}{m} \hat{r}_{ij}\hat{r}_{ij} \cdot
(\vec{p}_j-\vec{p}_i) \;\;\; \mbox{and} \;\;\; \dot{q}_{ij}^D =
\lambda(r_{ij}) \left(\frac{1}{\theta_i} -\frac{1}{\theta_j}\right)
        \label{7}
\end{equation}
where $\zeta(r_{ij})$ and $\lambda(r_{ij})$ are arbitrary positive
coefficients, that are even functions under 
time-reversal\cite{JBA}. A convenient choice
is to assume them to be functions of the slow variable
$r_{ij}$ only. Galilean invariance of $\vec{F}_{ij}^D$ and
$\dot{q}_{ij}^D$ is thus guaranteed. In this way, all the deterministic
interactions are completely specified.

        The properties of the random terms are chosen to also parallel the
theory of hydrodynamic fluctuations\cite{vSa,Lan}. Since $\vec{F}_{ij}^D$ and
$\dot{q}_{ij}^D$ are not coupled, we will demand that the random terms
$\vec{F}_{ij}^R$ and $\dot{q}_{ij}^R$ be statistically independent. They
can be written in the form
\begin{equation}
\vec{F}_{ij}^R = \hat{r}_{ij} \, \Gamma_{ij} \, {\cal F}_{ij}(t)  \;\;\;
\mbox{and} \;\;\; \dot{q}_{ij}^R = \mbox{Sign}(i-j) \, \Lambda_{ij} \, {\cal
Q}_{ij}(t)         \label{8}
\end{equation}
where the function Sign$(i-j)$ is $1$ if $i>j$ and $-1$ if $i<j$, ensuring
that $\dot{q}_{ji}^R = -\dot{q}_{ij}^R$.
The scalar random variables ${\cal F}_{ij}$ and ${\cal Q}_{ij}$ are
stationary, Gaussian and white\cite{Maz2,JBA}, with zero mean and
correlations 
\begin{equation}
\langle {\cal F}_{ij}(t) {\cal F}_{kl}(t') \rangle = \langle {\cal Q}_{ij}
(t) {\cal Q}_{kl} (t') \rangle = (\delta_{ik}\delta_{jl}
+\delta_{il}\delta_{jk} ) \, \delta (t-t')     \label{9}
\end{equation}
$\Gamma_{ij}$ and $\Lambda_{ij}$ are functions to be determined later.
Note that $\zeta$ and $\Gamma_{ij}$ are, respectively, $\gamma \omega_D$ and
$\sigma \omega_R$ in ref.\cite{Esp1}.

        The stochastic differential equations eqs. (\ref{4}), (\ref{5}) and
(\ref{6}), together with eqs. (\ref{7}) and the properties of
the random terms given in eqs. (\ref{8}) and (\ref{9}), lead to the
Fokker-Planck equation 
\begin{equation}
\frac{\partial}{\partial t} P(\{\vec{r}_i\},\{\vec{p}_i\},\{u_i\})= L^{con}
P(\{\vec{r}_i\},\{\vec{p}_i\},\{u_i\}) + L^{dif}
P(\{\vec{r}_i\},\{\vec{p}_i\},\{u_i\})          \label{10}
\end{equation}
where the {\em convective} operator, $L^{con}$, and the {\em diffusive}
operator, $L^{dif}$, are defined as
\begin{eqnarray}
L^{con} & \equiv & -\sum_{i=1}^N \left[\frac{\partial}{\partial \vec{r}_i} \cdot
\frac{\vec{p}_i}{m} +\frac{\partial}{\partial \vec{p}_i} \cdot
\vec{F}_i^{ext} \right]-\sum_{i,j\neq i}^N \left\{ \frac{\partial}{\partial
\vec{p}_i} \cdot \left[ \vec{F}_{ij}^C + \zeta(r_{ij}) \frac{1}{m}
\hat{r}_{ij}\hat{r}_{ij} \cdot (\vec{p}_j-\vec{p}_i) \right] + \right. 
\nonumber \\
& + & \left. \frac{\partial}{\partial u_i} \left[\frac{1}{2m^2}\zeta(r_{ij})
\left[(\vec{p}_j-\vec{p}_i)\cdot \hat{r}_{ij} \right]^2 + \lambda(r_{ij})
\left(\frac{1}{\theta_i} -\frac{1}{\theta_j}\right) \right] \right\}
\label{11} \\
L^{dif} & \equiv & \sum_{i,j\neq i}^N \left\{ \frac{\partial}{\partial
\vec{p}_i} \cdot \frac{1}{2} \Gamma_{ij}^2 \hat{r}_{ij}\hat{r}_{ij} \cdot
\vec{\cal L}_{ij} + \right. \nonumber \\
&+& \left. \frac{\partial}{\partial u_i} \left[
\frac{1}{2m}(\vec{p}_j-\vec{p}_i) \cdot \frac{1}{2} \Gamma_{ij}^2
\hat{r}_{ij}\hat{r}_{ij} \cdot \vec{\cal L}_{ij} + \frac{1}{2}
\Lambda_{ij}^2 \left(\frac{\partial}{\partial u_i}-\frac{\partial}{\partial
u_j} \right) \right]\right\}  \nonumber \\
 & &  \label{12}
\end{eqnarray}
where we have in addition defined the operator
\begin{equation}
\vec{\cal L}_{ij} \equiv \left(\frac{\partial}{\partial \vec{p}_i}
-\frac{\partial}{\partial \vec{p}_j} \right) +\frac{1}{2m}
(\vec{p}_j-\vec{p}_i) \left(\frac{\partial}{\partial
u_i}+\frac{\partial}{\partial 
u_j} \right) \label{13}
\end{equation}
Imposing that the equilibrium distribution function given in eq. (\ref{Pe}) 
is a stationary solution of eq. (\ref{10}) \footnote{In fact, the
probability flux must also vanish for a system in thermodynamic
equilibrium\cite{JBA}}, we derive the corresponding
fluctuation-dissipation theorems
\begin{eqnarray}
\Gamma_{ij}^2 &=& 2 k \, \Theta_{ij} \, \zeta_{ij} \label{14} \\
\Lambda_{ij}^2 &=& 2 k \, \lambda_{ij}         \label{14b}
\end{eqnarray}
where we have defined the mean inverse temperature as $\Theta_{ij}^{-1}
=(1/\theta_i+1/\theta_j)/2$, that is a function only of $u_i$ and $u_j$.
Eq. (\ref{10}) together with eqs.
(\ref{14}) and (\ref{14b}) are a generalization of the results found in
ref.\cite{Esp1}, which refer to isothermal conditions and momentum
conservation only. A crucial
difference 
regarding the momentum change is, however, the fact that 
the fluctuation-dissipation theorem given in eq. (\ref{14}) relates the
strength of the random force with the temperature $\Theta_{ij}$,
instead of the thermodynamic temperature $T$\cite{Esp1}. Thus, the model
presented here is defined in 
terms of particle properties only, with no reference to macroscopic
magnitudes such as the thermodynamic temperature or the density of the
system. 
This neat property permits the use of DPD in situations other than
isothermal. Together with the updating algorithms shown below and the
second fluctuation-dissipation, eq. (\ref{14b}), this is the
main result of this letter.

        Note that the Langevin equations eqs. (\ref{5}) and (\ref{6}) are
subject to the so-called It\^o-Stratonovich dilemma\cite{vK} due to the
occurence of products of fast variables, such as $\vec{p}_i$ or
$\Theta_{ij}$, and random variables which are $\delta$-correlated in time.
We take, 
however, eq. (\ref{10}) as the true definition of the random
processes.  The choice 
made here for these random processes is based on the {\em detailed balance
condition} and on the {\em weak coupling assumption}\cite{JBA}, which permit
that the properties of the random forces be independent of
overall thermodynamic properties. Finally, the updating
algorithms can directly be obtained from eq. (\ref{10}), using eqs.
(\ref{14}) and (\ref{14b}), with no ambiguity.
After some algebra, we obtain the new values of the variables $\vec{r}~'_i
\equiv \vec{r}_i(t+\delta t)$, $\vec{p}~'_i
\equiv \vec{p}_i(t+\delta t)$ and $u_i' \equiv u_i (t+\delta t)$ in terms
of the old ones at $t$
\begin{eqnarray}
\vec{r}~'_i &=& \vec{r}_i + \frac{\vec{p}_i}{m} \delta t       \label{15a} \\
\vec{p}~'_i &=& \vec{p}_i +\left\{\vec{F}_{i}^{ext} + \sum_{j\neq i}
\left[\vec{F}_{ij}^C + \frac{1}{m} \xi \,
(\vec{p}_j-\vec{p}_i) \cdot \hat{r}_{ij} \hat{r}_{ij} \right] \right\}
\delta t + \nonumber \\
 &+& \sum_{j\neq i} \hat{r}_{ij} \sqrt{2 k \Theta_{ij} \zeta(r_{ij}) \, \delta
t} \; \Omega_{ij}^{(p)} \label{15} \\ 
u_i' &=& u_i + \sum_{j \neq i} \left\{\frac{1}{2m^2}
\xi
\left[(\vec{p}_j-\vec{p}_i) \cdot \hat{r}_{ij} \right]^2 + \lambda
(r_{ij}) \left(\frac{1}{\theta_i}-\frac{1}{\theta_j} \right) -\frac{1}{m}
k \, \Theta_{ij} \zeta (r_{ij}) \right\} \delta t + \nonumber \\ 
&+& \sum_{j \neq i} \left \{ \frac{1}{2m} 
(\vec{p}_j-\vec{p}_i) \cdot \hat{r}_{ij}\sqrt{2 k \Theta_{ij}
\zeta(r_{ij}) \, \delta t} \; \Omega_{ij}^{(p)} + \sqrt{2 k \lambda (r_{ij})
\, \delta t} \; \Omega_{ij}^{(q)} \right\}       \label{16}
\end{eqnarray}
where we have defined an effective friction coefficient
\begin{equation}
\xi(r_{ij},u_i,u_j) \equiv \zeta(r_{ij}) \left[ 1 +
\frac{k}{2}\left(\frac{\partial}{\partial u_i} + \frac{\partial}{\partial
u_j} \right) \Theta_{ij} \right]. \label{17}
\end{equation}
$\Omega_{ij}^{(p)}$ and $\Omega_{ij}^{(q)}$ are independent random
numbers defined from the random variables ${\cal F}_{ij}$ and ${\cal
Q}_{ij}$ and are Gaussian with zero mean and variance $\langle
\Omega_{ij}^{(p)} \Omega_{kl}^{(p)}\rangle = \langle
\Omega_{ij}^{(q)} \Omega_{kl}^{(q)}\rangle = (\delta_{ik}\delta_{jl} +
\delta_{il} \delta_{jk} )$.

        The model presented in this letter constitutes a 
generalization of the DPD method. The addition of energy conservation in
the particle-particle interaction in a consistent way, allows the
derivation of an updating algorithm defined in terms of particle variables
only. We have obtained two fluctuation-dissipation theorems which permit
that the proper thermodynamic equilibrium be reached for the model under
appropriate conditions. It should be mentioned that, with respect to
previous treatments\cite{Esp1}, the fluctuation-dissipation theorem
for the random force derived in this work contains the local temperature
$\Theta_{ij}$ instead of the 
thermodynamic temperature $T$, stressing the local nature of our model. In
addition, the fluctuation-dissipation theorem for the heat flux has no
counterpart in previous DPD formulations. The corresponding computer
implementation of the model has 
already shown that the total energy is well conserved for an isolated
system, although some dependency on the time step $\delta t$ has been
observed. Therefore, our model can sustain temperature gradients and thus
heat flow can be described.


\begin{thebibliography}{99}

\bibitem{Hoo} Hoogerbrugge, P.J. and Koelman, J.M.V.A, Europhys. Lett.,
{\bf 19} (1992) 155.

\bibitem{Esp1} Espa\~nol, P., Warren, P., Europhys. Lett., {\bf 30} (1995)
191. 

\bibitem{Esp2} Espa\~nol, P., Phys. Rev. E, {\bf 52} (1995) 1734.

\bibitem{Ern} Marsh, C., Backx, G. and Ernst, M.H. (submitted to Phys.
Rev. E).

\bibitem{Lek} Boek, E.S., Coveney, P.V., Lekkerkerker,
H.N.W. and van der Schoot, P., Phys. Rev. E (to appear). 

\bibitem{Cov} Coveney, P.V. and Novik, K.E., Phys. Rev. E, {\bf 54} (1996)
5134.

\bibitem{Maz1} de Groot, S.R. and Mazur, P., {\em Non-equilibrium
thermodynamics}, (Dover, New York) 1984.

\bibitem{vSa} van Saarloos, W., Bedeaux, D. and Mazur, P., Physica {\bf
110A} (1982) 147. 

\bibitem{Lan} Landau, L.D. and Lifshitz, E., {\em Fluid Mechanics},
(Pergamon, New York) 1984.

\bibitem{Ca} Callen, H.B., {\em Thermodynamics and an Introduction to
Thermostatistics}, (John Wiley \& Sons, New York) 1985.

\bibitem{JBA} Avalos, J.B. and Pagonabarraga, I., Phys. Rev. E {\bf 52}
(1995) 5881.

\bibitem{Maz2} Mazur, P. and Bedeaux, D., Physica {\bf 173A} (1991) 155.

\bibitem{vK} van Kampen, N.G., {Stochastic Processes in Physics and
Chemistry}, 2nd. edition, (North Holland, Amsterdam) 1992.

\end{thebibliography}
\end{document}